# Electric field control of magnetocaloric effect in cylindrical MnAs/PZT magnetoelectric composite

Abdulkarim A. Amirov[1]*, Maksim A. Koliushenkov[2], Abdula A. Mukhuchev[1], Dibir M. Yusupov[1], Valeriya V. Govorina[3], Dmitriy S. Neznakhin[3], Gennady A. Govor[4], Akhmed M. Aliev[1]

[1]Amirkhanov Institute of Physics of Dagestan Federal Research Center, Russian Academy of Sciences, 367003, Makhachkala, Russia

[2]Physics Department, Lomonosov Moscow State University, 119991 Moscow, Russia

[3]Ural Federal University, 620002 Ekaterinburg, Russia

[4]Scientific-Practical Materials Research Center of National Academy of Sciences of Belarus, 220072 Minsk, Belarus

* Corresponding author, E-mail: *amiroff_a@mail.ru*

## Abstract

The possibility of electric field control of magnetocaloric effect through quasi-isostatic compression as a result of the converse piezoelectric effect was demonstrated on cylindrical type magnetoelectric composite MnAs/PZT. It was shown that an electric voltage of 100 V corresponding to an electric field of E ~0.3 kV/mm applied to the walls of the piezoelectric component PZT of the MnAs/PZT composite contributes to an increase in the maximum adiabatic temperature change by 0.2 K in the temperature range of the magnetostructural phase transition of MnAs ~317 K at magnetic field change of 1.8 T. Calculations using the finite element method have shown that an electric field voltage of 100 V is capable of creating a quasi-isostatic mechanical stress in the region inside a cylindrical PZT tube of ~3 MPa. Moreover, in the region of weak pressures up to 10 MPa, the contribution to the MCE from piezo compression linearly depends on the electrical voltage that can be used for control the MCE.

# 1.Introduction.

One of the important tasks in the study of multicaloric material is connected to control of caloric (magnetocaloric (MCE), electrocaloric (ECE), mechanocaloric (MechCE)) effects trough external fields (magnetic, electric, elastic) with differ nature [1,2]. Multicaloric composites with magnetoelectric (ME) interaction and different types of connectivity are one of the prospective type of these materials. [3]. The most of well-known ME composites have layered planar structure, consisted of piezoelectric and magnetostrictive phases and referred to 2-2 type of connectivity. The degree of the ME interaction is typically evaluated separately through the transverse and longitudinal ME coefficients $\alpha_{33}$ and $\alpha_{31}$ respectively. These interactions are related to the piezoelectric modules $d_{33}$ and $d_{31}$, which characterize the longitudinal and transverse piezoelectric effects, respectively. Unlike layered composites, in cylindrical composites, due to their geometry, the magnetostrictive and piezoelectric components are mechanically connected better. The piezoelectric coefficients $d_{33}$ and $d_{31}$ simultaneously contribute to the ME interaction, which can eventually lead to the boosting of ME effect [4,5].

The model of multicaloric composite proposed in our work is referred to 1-3 type connectivity, where the piezoelectric component has a thin tube-like shape and inner part of this tube is filled with a magnetocaloric material. In this case, the ME coupling is facilitated by the interaction of a magnetostrictive (material with large magnetostriction and «giant» MCE) and a piezoelectric (material with high piezoelectric coefficients $d_{33}$ and $d_{31}$) components. Inorganic ME composites of type 1-3, in contrast to type 0-3 and 2-2, are less prevalent and are primarily represented by two- or three-layer cylindrical composites. These can be conceptualized as one or more coaxially inserted alternating components of magnetostrictive and piezoelectric materials [6]. For example, in [7], a cylindrical three-layer Ni-PZT-Ni ME composite was considered as a "twisted" type 2-2 three-layer composite. Furthermore, studies examining cylindrical ME composites frequently employ a ring configuration, wherein the diameter of the ring composite $d$ is greater than its length $l$, that is, $d > l$, and ME interaction is typically evaluated through the $d_{31}$ component of the piezoelectric coefficient (transverse piezoelectric effect), with the applied electrical voltage oriented perpendicular to the direction of measurement of the piezoelectric displacement [4]. In contrast to the samples previously described, the model of our proposed multicaloric composite of connectivity type 1-3 has the geometry of a thin cylinder with a length much greater than its diameter ($l >> d$). The cylinder's cavity is filled with a magnetocaloric material which acts as a magnetostrictive component (Figure 1). In the model under consideration, there is no mechanical action in the direction along the axis of the cylinder. Instead, the interaction between the composite components is realized only through the inner surface of the tube, which is mechanically connected with the magnetostrictive component that fills the inner cavity of the tube.

Electrical contacts are applied to the inner and outer surfaces of the piezoelectric tube according scheme described in Figure 1a.

As known, one of the approaches to controlling MCE is the application of mechanical pressure, which can be realized in combination with the application of a magnetic field. This can be either uniaxial compression (or stretching) [8] or isostatic compression [9]. This approach has been demonstrated for controlling by parameters of phase transition, magnetic and magnetocaloric properties in materials with first order magnetic phase transition (FOMPT), including La-Fe-Si [10,11], MnAs [9], and FeRh [12], where the hydrostatic pressure is employed as an additional external stimulus to induce the isostatic compression.

The proposed idea is based on the concept of MCE control through quasi-isostatic piezoelectric compression, when the functional parameters (piezoelectric constants, operated temperature range and etc.) of piezoelectric ceramics and magnetocaloric parameters (maximum of adiabatic temperature changes, temperature of maximum of adiabatic temperature changes and etc.) of magnetic components are specially selected. In other words, the objective of the present study is to investigate the prospective for manipulation the magnetic and magnetocaloric properties of a FOMPT materials via quasi-isostatic compression induced by the inverse piezoelectric effect. The novelty of the proposed approach is that, in our specific context, the medium of impact transmission is not a high-pressure liquid, but a piezoelectric material. In order to realize the proposed concept, we have chosen to employ lead zirconate titanate (PZT) as the piezoelectric component and MnAs as a magnetic component; as material with «giant» MCE at the room temperature region [13,14].

**2.Materials and Methods.**

A sample of a cylindrical MnAs/PZT composite consisted was fabricated by packing powder of MnAs magnetocaloric material into a thin ceramic tube made of commercial piezoelectric material of PZT (ELPA LLC). According to the manufacturer's data, the PZT has piezoelectric coefficients as follows: $d_{31} = -170 \cdot 10^{-12}$ C/N and $d_{33} = 350 \cdot 10^{-12}$ C/N. A sample of manganese arsenide (MnAs) single crystal obtained via the Bridgman–Stockbarger method was utilized as the initial sample for the magnetic component of the composite [15].

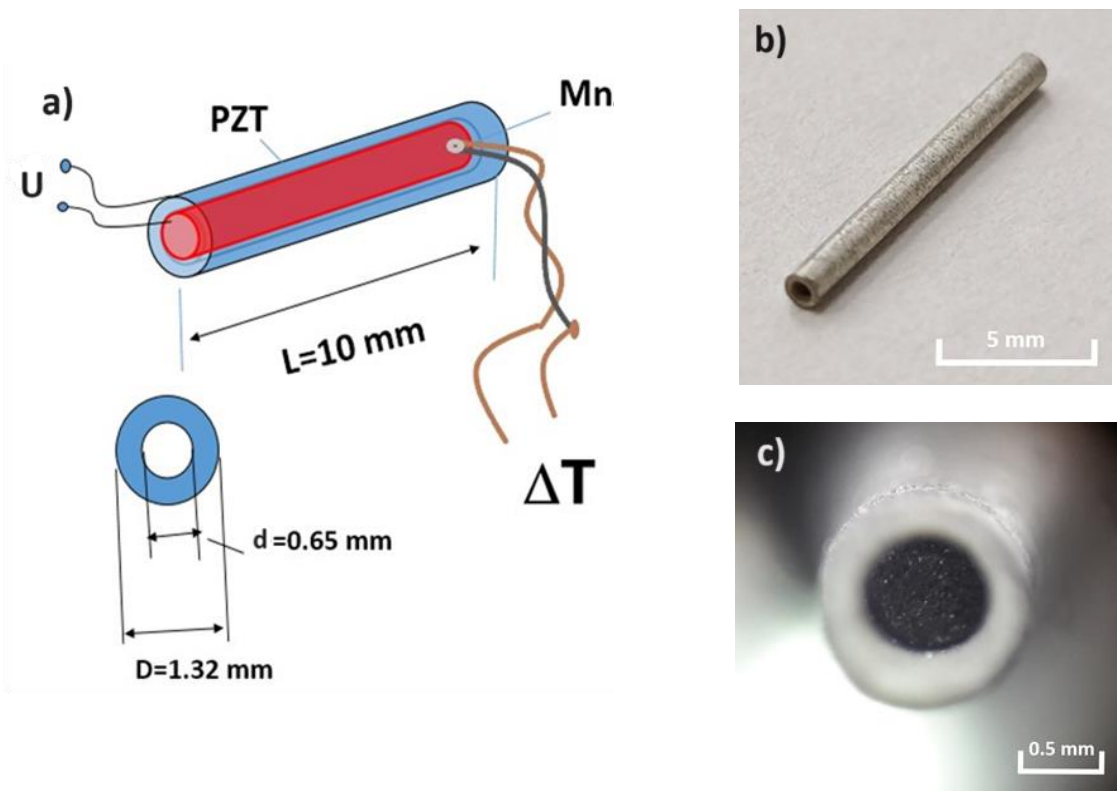

*Figure 1. Model of the multicaloric MnAs/PZT composite (a), photographs of the PZT tube used to fabricate the composite (b) and one of the bases of the cylindrical MnAs/PZT composite (c).*

For the fabrication of the composite, the MnAs sample was milled in a ceramic mortar to obtain a granules with sizes less than 200 μm. The structure of MnAs was confirmed by X-ray diffraction analysis.

A two-component epoxy composition was employed as a binder for packing the powder into a cylindrical PZT tube. The final mixture was then tightly tamped into a PZT ceramic tube with a length of 10 mm, and outer and inner diameters of 1.32 and 0.65 mm, respectively; this tube had been pre-insulated on one side (Figure 1b-c). The samples were dried at 18°C for 48 hours, which is below the Curie temperature for both ferroelectric and magnetic components of MnAs/PZT composite. The mass fraction of the adhesive binder after the drying process was approximately 5% of the total mass of MnAs powder. The microstructure of the samples was investigated by optical and scanning electron microscopy. Magnetic properties were studied using vibration magnetometry, thermal expansion and magnetostriction were measured by strain gauge method. In order to better explanation of the nature of magnetostructural phenomena, magnetization and magnetostriction were measured in simultaneous mode using a special experimental insert, described elsewhere [11]. The MCE was investigated using direct method, by measurements of the adiabatic temperature change $\Delta T_{AD}$ when an alternating magnetic field was applied, in accordance with the methodology described in [16]. Two modes of adiabatic temperature change measurement were employed: "*switch on*", in which an electric voltage was applied to the piezoelectric tube in accordance with the scheme depicted in Figure 1, and "*switch off*", in which the applied voltage was set to zero. The applied electric voltage was 100 V, which corresponds to an electric field of $E=U/t\sim300$ *V/mm* at a piezoelectric tube wall thickness $t=(D-d)/2$=0.335 mm. The electrical contacts were fabricated by applying a silver paste to the outer and inner sides of the piezoelectric tube (Figure 1, b-c). To obtain more precise measurements of the adiabatic temperature change, a copper-constantan thermocouple with 25 μm diameter wires, which were pre-twisted and glued, was inserted into the tube (Figure 1a). As magnetic field source for MCE measurements the adjustable magnetic system with a Halbach-type structure and a magnetic field amplitude value of 1.8 T.

### 3. Results and Discussion.

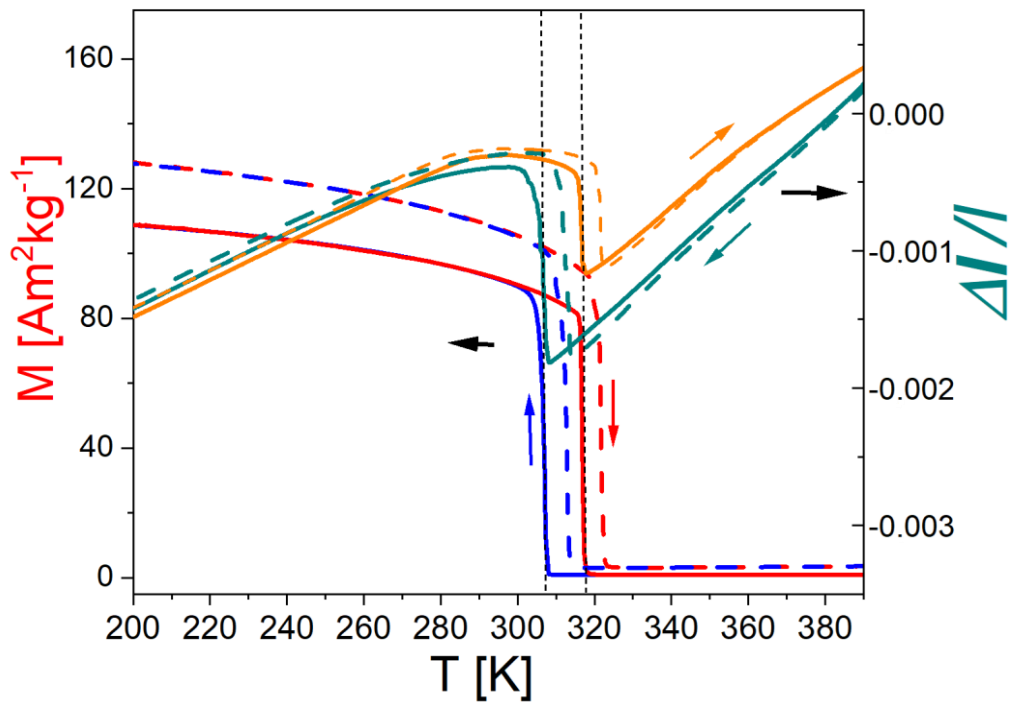

*Figure 2. Temperature dependences of magnetization and the relative change in linear dimensions Δl/l in magnetic fields of 0.5 T (solid line) and 2 T (dashed line) for the MnAs single crystal utilized for the fabrication of the MnAs/PZT composite magnetic component.*

Figure 2 show the temperature dependencies of the magnetization and the relative change in linear dimensions in magnetic fields of 0.5 T and 2 T of a MnAs sample, measured in heating and cooling modes. The abrupt change in magnetization observed in the region of 320 K is indicative of a FOMPT from the ferromagnetic state to the paramagnetic state. This transformation involves the structural transition of the hexagonal structure of the B8$_1$ type (α-phase, symmetry group P6$_3$/mmc) into the orthorhombic structure of the B3$_1$ type (β-phase, symmetry group Pnma) [15]. Observed FOMPT is accompanied by giant changes in lattice volume (0.17%), significant magnetostriction (~2%), and the release of a considerable latent transition heat [17–19]. According to [17,20], at the magnetostructural phase transition from ferromagnetic (B8$_1$) to paramagnetic (B3$_1$), there is a decrease in the lattice volume, which is reflected in the behavior of the temperature dependence of the relative linear expansion, which shows a minimum in the transition temperature region (Figure 2).

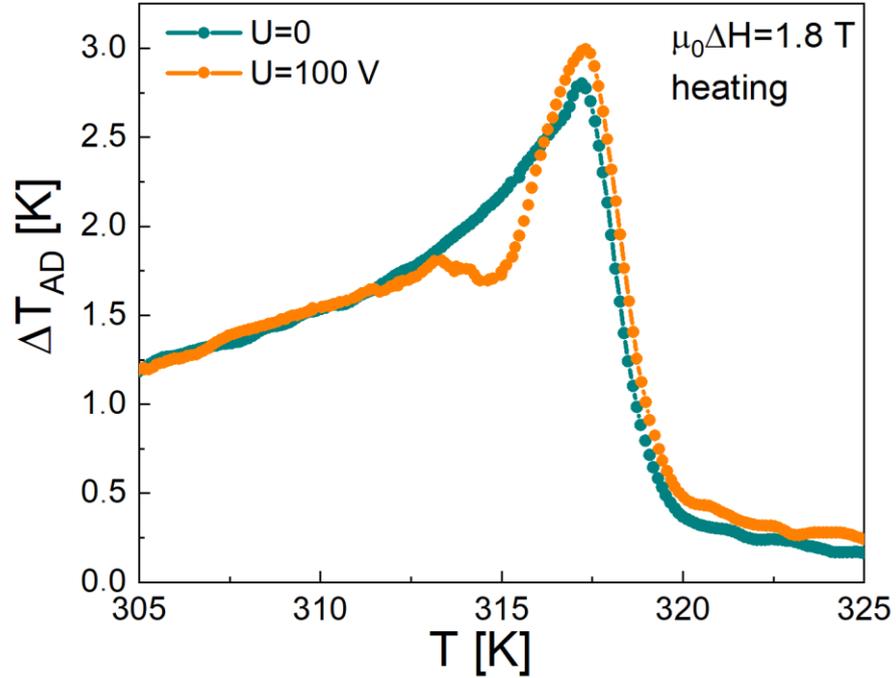

*Figure 3. Temperature dependences of the adiabatic temperature change ΔT of the multicaloric MnAs/PZT composite sample measured by the heating protocol under the application of a 1.8 T magnetic field in the "switch on" and "switch off" modes.*

Figure 3 shows the temperature dependencies of the adiabatic temperature change ΔT$_{AD}$ measured in the heating protocol under the application of a 1.8 T magnetic field in «*switch on*" and "*switch off*" modes. In the absence of an electric voltage on the cylindrical PZT matrix, the MCE in the sample exhibits behavior characteristic of a bulk MnAs sample, with a maximum of 2.8 K at 317.2 K. The application of 100 V of electric voltage results in a minor alteration to the ΔT$_{AD}$(T) dependence, exhibiting a slight increase in the value of ΔT$_{AD}$ maximum up to 3 K at 317.3 K. The enhancement of MCE from 47 to 267 J*kg$^{-1}$*K$^{-1}$ due to hydrostatic compression by applying a pressure of 223 MPa and a magnetic field of 5 T was observed in [9].

As known in FOMPT materials the total entropy of the system $S_T$ as a function of temperature $T$, pressure $P$ and magnetic field $\mu_0 H$ is mainly characterized by contributions of the magnetic $S_{mag}$ and lattice $S_{latt}$ subsystems, when

$$S_T(T, P, \mu_0 H) = S_{mag} + S_{latt} \qquad (1).$$

Other contributions to the total entropy of the system are neglected because they are small. To determine the upper limit of the entropy change, expression

$$\Delta S^{max} = R\,(2J+1),\qquad(2)$$

is often used, based on the hypothesis that the lattice and electron contributions to the total entropy are independent of the magnetic field. However, as experiments and calculations show that "giant" MCE is observed in MnAs under pressure, the magnitude of the maximum $\Delta S = 267$ J*kg$^{-1}$*K$^{-1}$ (5 T and 2.23 kbar) significantly exceeds the calculated upper limit $\Delta S_{max} = 103$ J*kg$^{-1}$*K$^{-1}$[9,21,22]. The strong magnetoelastic interaction resulting from the combined action of a magnetic field and pressure in MnAs serves as the driving force for a FOMPT. Furthermore, it makes a significant contribution to the lattice entropy, thereby influencing the total caloric effect. The preliminary results obtained from a phenomenological model based on the development of the Bean-Rodbell model applied to pressurized MnAs confirm that the lattice contribution can explain the observed increase in MCE [21].

An original approach to estimate the magnetic and structural contributions to the overall MCE for MnAs based experiments of direct measurements of adiabatic temperature change and magnetostriction was proposed in [14]. The authors have demonstrated that in MnAs, the predominant contribution to the total MCE is provided by the structural subsystem, which accounts for nearly 70% of the total. Thus, one of the strategies for control of MCE can be considered to be the impact on the structural subsystem, including mechanical stimuli. As can be observed, the MCE of MnAs is responsive to even small pressures. The mechanical effect created in our experiment as a result of the inverse piezoelectric effect, taking into account the cylindrical shape of the sample, was initially assumed to be analogous to the action of hydrostatic pressure, at which isostatic compression occurs. However, as we know, the phase transition temperature in MnAs is one of the most sensitive to the pressure ($dT_c/dp = -165$ K*GPa$^{-1}$), and shifts to low temperatures, while in our case the MnAs transition temperature practically does not change under the effect of piezoelectric compression. Obviously, the key question in such a case is the real nature of the deformations and the magnitudes of the mechanical stresses that act on the MnAs component of the cylindrical MnAs/PZT composite sample when an electric voltage is applied to it according to the scheme shown in Figure 1. To answer this question, we performed calculations using the COMSOL Multiphysics package on the composite model with configuration and physical parameters corresponding to the real sample. Table 1 shows the main parameters of the MnAs composite components used for modeling.

*Table 1. Physical parameters of MnAs and PZT for mechanical strain calculations in MnAs/PZT composite, where $\rho$ is density, $Y$ is Young's modulus, $d_{33}$ is piezoelectric coefficient (in 33 mode), $T_C$ is Curie temperature for magnetic and ferroelectric components, $\varepsilon^\sigma_{33}/\varepsilon_0$ - relative permittivity.*

| Material | Symbol | $\rho$, kg/m³ | Y, GPa | $d_{33}$, pC/N | $T_C$, K | $\varepsilon^T_{33}/\varepsilon_0$ | Ref. |
|---|---|---|---|---|---|---|---|
| Pb(Ti$_{1-x}$Zr$_x$)O$_3$ | PZT | 7500 | 70 | 350 | 563 | 1750 | Manufacturer's data |
| MnAs | MnAs | 6310 | 32,2 | - | 17 | - | [13,19] |

Figure 4 shows the strain distribution along the radial axis of an empty PZT piezoelectric tube and one filled with the magnetocaloric material MnAs, corresponding to the case of the MnAs/PZT composite.

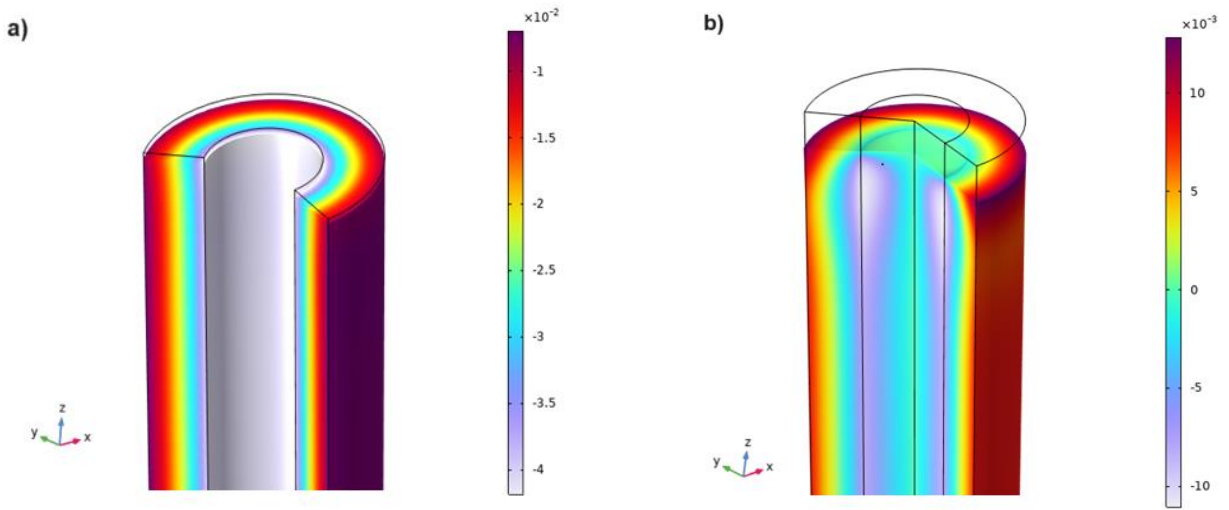

*Figure 4. Strain distribution (μm) along the radial axis of the empty PZT piezoelectric tube and the MnAs/PZT composite when a voltage of 100 V is applied.*

The maximum deformation is observed on the inside of the empty piezoelectric tube and is about 0.04 μm, while in the case of the MnAs/PZT composite, the presence of elastic coupling between MnAs and PZT prevents PZT deformation. The pressure induced in MnAs by the reverse piezoelectric effect, when an electric voltage of 100 V is applied, is uniformly distributed, except for the edge regions and is ~3 MPa (Figure 5).

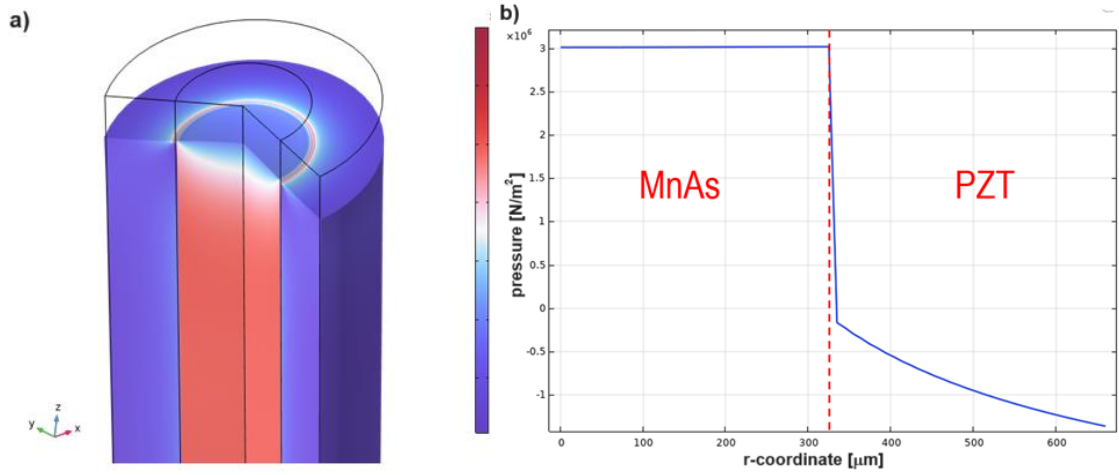

*Figure 5. 3D map of pressure (Pa) distribution over the volume (a) and along the radial axis of the MnAs/PZT composite.*

The region of inhomogeneous mechanical pressure distribution along the axis of the sample occupies about 1 mm in length from each of the edges of the piezoelectric tube, with the volume of inhomogeneous pressure distribution accounting for about 20% of the total volume of the composite. According to [9] for MnAs in a 5 T magnetic field, application of a hydrostatic pressure of 38 MPa leads to an increase in magnetic entropy by ~60 J*kg*K$^{-1}$, which corresponds to a change of ~6 J*kg*K$^{-1}$ per 3 MPa. This indirectly confirms our observed results on the increase of MCE due to piezocompression of ~3 MPa. The fact that the transition temperature is almost unchanged by the application of pressure in our case may be related to the inhomogeneous nature of deformation at the edges of the MnAs/PZT composite. It can be assumed that the compression induced by the piezo effect stimulates the MnAs volume change in the region of the first-order magneto-structural phase transition, where the volume of the MnAs single crystal sample decreases anomalously according to the measurements of the relative change of the linear dimensions $\Delta l/l$ in magnetic fields of 0.5 and 2 T (Figure 2).

If the piezoelectric tube is considered as an unfolded rectangular PZT plate, the mechanical action induced by the inverse piezo effect can be considered analogous to the problem of a rectangular piezoactuator consisting of a set of piezoelectric plates N = 1 and can be described by a general equation:

$$S = s^E \sigma + d_{33} E, \qquad (3)$$

where $S=\Delta l/l$ is the relative displacement, and $s^E=I/Y$ is the stiffness coefficient inversely proportional to the Young's modulus $Y$ in the specified direction (in this case, $Y=Y_{33}$). In consideration of the fact that mechanical stress inhibits deformation due to the piezo effect, equation (3) will assume the following form:

$$S = -s^E + d_{33} E. \qquad (4)$$

Equation (4) can be represented in the following form:

$$\frac{\Delta l}{l_0} Y_{33} S_0 = d_{33} Y_{33} S_0 E - S\sigma. \qquad (5)$$

The left-hand side of equation (5) pertains to the elastic deformation force

$$F_{elast} = \frac{\Delta l}{l_0} Y_{33} S_{0_{33}}, \qquad (6)$$

first summand on the right-hand side of equation (5)

$$F_{electr} = d_{33} Y_{33} S_0 E = \frac{d_{33} Y_{33} S_0 U}{t} \qquad (7)$$

describes the mechanical force exerted by the electric field, and the second summand on the right-hand side of equation (5) describes the static force applied to the magnetic component of MnAs with mass $m_0$. In the present experiment, the static loading was employed, and the dynamic forces resulting from the motion of the object undergoing the action were not considered. As depicted in equation (7), if we assume that within the specified temperature range, $d_{33}$ and $Y_{33}$ exhibit minimal variation, the mechanical force generated by the application of the electric field is directly proportional to the electric voltage $U$, which can be exploited to regulate the pressure generated inside the piezoelectric tube.

The total adiabatic temperature change $\Delta T_{AD}^{total}$ in the MnAs/PZT composite resulting from the combination of an applied magnetic field and piezoelectric compression can be represented as the sum of the contributions from net MCE $\Delta T_{AD}^{mag}(H)$ and ME interaction $\Delta T_{AD}^{ME}(E)$, which are functions of magnetic and electric fields, respectively:

$$\Delta T_{AD}^{total} = \Delta T_{AD}^{mag}(H) + \Delta T_{AD}^{ME}(E). \qquad (8)$$

The adiabatic temperature change due to MCE is found to be nonlinearly dependent on the magnetic field, with a dependence of $\Delta T \sim \Delta H^n$, where $n = 0.66$ was observed in high magnetic fields above ~3 T [14]. In our experiments the applied magnetic field did not exceed 1.8 T, and the character of the magnetofield dependence $\Delta T_{AD}^{mag}$ was found to be approximately linear. As for the second term in equation (8), the final stimuli is pressure, which is proportional to the electric voltage used to control as a result of the converse piezoelectric effect.

Figure 6 show the calculated and experimental dependencies of adiabatic temperature change in a 1.8 Tesla magnetic field as a function of electric voltage, plotted on the basis of equation (8). A linear field dependence of the type $y = kx + b$ is observed for $\Delta T_{AD}^{total}$, where coefficient $k = d(\Delta T)/d(\Delta U)$ is the change of $\Delta T$ by 1 V of applied electric voltage, and $b$ is the net MCE $\Delta T$ when the magnetic field changes by 1.8 T in a zero electric field. For comparison, on Figure 6 is also illustrated the dependence of the pressure induced in the MnAs/PZT composite on the controlling electric voltage, as obtained on the basis of calculations conducted using the COMSOL software. It is important to note that the given dependence of adiabatic temperature change in MnAs/PZT composite as a result of combination of applied magnetic field and piezoelectric compression, based on equation (8), is a simplified form that does not take into account hysteresis and nonlinear character of dependence of linear dimensions change as a result of inverse piezoelectric effect on the applied electric field. The nonlinear character is more pronounced in the region of voltages in the vicinity of 400 V and above. A further increase in voltage results in ohmic heating of the piezoelectric material and its electrical breakdown. Nevertheless, equation (8) can be employed in the area of stimulation utilizing modest loads and control voltages (less than 100 V) and a temperature range proximate to room temperature, wherein the piezoelectric characteristics of PZT ceramics are relatively stable. Additionally, the equation delineating the correlation between MCE and the electric voltage inducing piezoelectric compression in MnAs/PZT composites can be extended, taking into account the aforementioned factors.

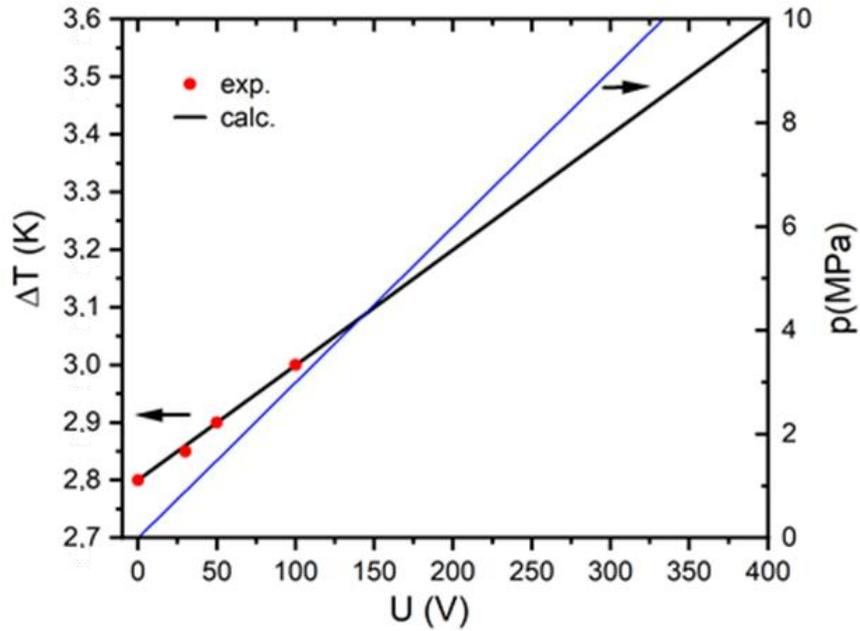

*Figure 6. The experimental and calculated dependences of the adiabatic temperature change of the MnAs/PZT composite in a 1.8 T magnetic field on the controlling electric voltage inducing piezoelectric compression (left) and the dependence of the pressure induced as a result of the inverse piezoelectric effect in the MnAs/PZT composite on the controlling electric voltage (right).*

Thus, on the basis of the obtained results of direct measurements of MCE under the action of quasi-isostatic piezostatic compression, calculation data and the results of literature data, we can conclude that even insignificant (up to ~3 MPa) pressures are able to influence the value of MCE by increasing the contribution to the structural subsystem of the total entropy change.